\documentclass{appolb}
\usepackage{epsfig}
\usepackage{subfig} % uses subfloat in figure environment
\usepackage{url}
\usepackage{wrapfig}
\usepackage{eufrak} % \EuFrak{name}
\usepackage{amsmath}   %for allowdisplaybreaks
\usepackage{microtype}\usepackage{lmodern}

\usepackage[T1]{fontenc}
\usepackage{eurosym}   % needed for Eurosymbols taken from ``Text'' \texteuro does not work

\usepackage{mathptmx}
\usepackage{graphicx}
\usepackage[normalem]{ulem} %emphasize further-on italic
\usepackage{color}
\definecolor{myred}{rgb}{0.6,0,0} %usage:  {\textcolor{myred}{Hello World}}
\definecolor{myblue}{rgb}{0,0.2,0.4}
\definecolor{mygreen}{rgb}{0,0.9,0.1}
\definecolor{Orange}{rgb}{1.,0.65,0.}
\usepackage{color}
\definecolor{myred}{rgb}{1.0,0,0} %usage:  {\textcolor{myred}{Hello World}}
\definecolor{mygreen}{rgb}{0,0.9,0.1} %usage:  {\textcolor{myred}{Hello World}}
\definecolor{myblue}{rgb}{0,0.2,0.4}
\definecolor{mygray}{rgb}{.8,.8,.8}
\definecolor{darkorange}{rgb}{1, 0.549, 0}
\definecolor{purple}{rgb}{0.6,0.4,0.6}
\definecolor{mymagenta}{rgb}{0.6,0.4,0.6}
 \definecolor{LightCyan} {rgb}{0.88,1.,1.}
 \definecolor{Orange} {rgb}{1.,0.65,0.}
 \definecolor{PaleGreen} {rgb}{0.6,0.98,0.6}
 \definecolor{Pink} {rgb}{1.,0.75,0.8}
\definecolor{Red}{rgb}{1,0,0}
   \definecolor{Blue}{rgb}{0,0,1}
   \definecolor{Yellow}{rgb}{1,1,0}
   \definecolor{Orange}{rgb}{1,0.4,0}
   \definecolor{Pink}{rgb}{1,0,1}
   \definecolor{Purple}{rgb}{0.5,0,0.5}
   \definecolor{Teal}{rgb}{0,0.5,0.5}
   \definecolor{Navy}{rgb}{0,0,0.5}
   \definecolor{Aqua}{rgb}{0,1,1}
   \definecolor{Lime}{rgb}{0,1,0}
   \definecolor{Green}{rgb}{0,0.5,0}
   \definecolor{Olive}{rgb}{0.5,0.5,0}
   \definecolor{Maroon}{rgb}{0.5,0,0}
   \definecolor{Brown}{rgb}{0.6,0.4,0.2}
   \definecolor{Black}{gray}{0}
   \definecolor{Gray}{gray}{0.5}
   \definecolor{Silver}{gray}{0.75}
   \definecolor{White}{gray}{1}
\usepackage{hyperref}  % should come last
\definecolor{darkblue}{rgb}{0,0,.5}
\hypersetup{
linktocpage=true,
    bookmarks=true,         % show bookmarks bar?
breaklinks=true,
    frenchlinks=false, %
    unicode=false,          % non-Latin characters in Acrobat??????s bookmarks
    pdftoolbar=true,        % show Acrobat??????s toolbar?
    pdfmenubar=true,        % show Acrobat??????s menu?
    pdffitwindow=false,     % window fit to page when opened
    pdfstartview={FitH},    % fits the width of the page to the window
    pdfauthor={J.Fleischer, T.Riemann},     % author
    pdfcreator={JF+TR},   % creator of the document
    pdfproducer={JF+TR}, % producer of the document
    pdfkeywords={}, % list of keywords
    pdfnewwindow=true,      % links in new window
    colorlinks=false,       % false: boxed links; true: colored links
    linkcolor=red,          % color of internal links
    citecolor=green,        % color of links to bibliography
    filecolor=magenta,      % color of file links
    urlcolor=cyan           % color of external links
}
\usepackage[all]{hypcap} % makes figures, tables etc. really visible when pointed to

\usepackage{rotating}
\newcommand{\be}{\begin{equation}}
\newcommand{\ee}{\end{equation}}
\newcommand{\bea}{\begin{eqnarray}}
\newcommand{\eea}{\end{eqnarray}}

\def \eps {\varepsilon}

\def \Yps {{\rm{Y}}}
\def \Dps {{\rm{D}}}
\newcommand{\nn}{\nonumber}

\begin{document}
\eqsec  % uncomment this line to get equations numbered by (sec.num)
\title{
\texttt{
\vspace*{-3cm}
\begin{flushleft}
DESY  11-206
\\
BI-TP 2011/37   %for the hep-ph of Ustron
\\
 SFB/CPP-11-66
\end{flushleft}
}
\vspace{1cm}
Simplifying $5$-point tensor reduction
\thanks{Presented at XXXV International Conference of Theoretical Physics
MATTER TO THE DEEPEST: Recent Developments in Physics of Fundamental Interactions, Ustron'11
}%
}
\author{J. Fleischer
\address{Fakult\"at f\"ur Physik, Universit\"at Bielefeld, Universit\"atsstr. 25,  33615
Bielefeld, Germany}
\and
 T. Riemann
\address{Deutsches Elektronen-Synchrotron, DESY, Platanenallee
  6, 15738 Zeuthen, Germany}
}
\maketitle
\begin{abstract}
The $5$-point tensors have the property that after insertion of the metric tensor $g^{\mu \nu}$ in terms 
of external momenta, all  $g^{\mu \nu}$-contributions in the tensor decomposition
cancel.  If furthermore the tensors are contracted with external momenta, the inverse
$5$-point Gram determinant $\left( \right)_5$ cancels automatically. If the remaining $4$-point
sub-Gram determinant ${s\choose s}_5$ is not small then this approach appears to be particularly
efficient in numerical calculations. We also indicate how to deal with small ${s\choose s}_5$.
Explicit formulae for tensors of degree $2$ and $3$ are given for large and small (sub-) Gram determinants.
\end{abstract}
  
% \begin{document}
\allowdisplaybreaks
\sloppy
%-----------------------------------

\section{\label{Intro} Introduction}
%======================================================================
In   \cite{Fleischer:2010sq}  we have worked out an algebraic method to present one-loop tensor integrals in terms of  scalar one-loop $1$-point to 
$4$-point functions.
The tensor integrals are defined as
\bea
\label{definition}
 I_{n}^{\mu_1\cdots\mu_R} &=&~\int \frac{d^dk}{i {\pi}^{d/2}}~~\frac{\prod_{r=1}^{R} k^{\mu_r}}{\prod_{j=1}^{n}c_j},
\eea
with denominators $c_j$,
\begin{eqnarray}\label{propagators}
c_j &=& (k-q_j)^2-m_j^2 +i \epsilon.
\end{eqnarray}
For the tensor decomposition we use Davydychev's approach~\cite{Davydychev:1991va}, recursion relations as
given in \cite{Fleischer:1999hq} and  make detailed use of \emph{modified Cayley determinants} introduced
for this purpose in \cite{Melrose:1965kb}. For these techniques and details of definitions we ask the reader to to consult
\cite{Fleischer:2010sq}. The following linear combinations of the chords 
have proven as particularly useful:
\bea
 Q_s^{\mu}&=&\sum_{i=1}^{5}  q_i^{\mu} \frac{{s\choose i}_5}{\left(  \right)_5},~~~ s=0 \cdots 5,
\label{Q6}
\\\label{3.11}
Q_s^{t,\mu}&=&\sum_{i=1}^{5} q_i^{\mu} \frac{ {st\choose it}_5}{{t\choose t}_5},~~~ s,t=1 \cdots 5.
\eea

\section{\label{Examples} Explicit examples}
%======================================================================
According to \cite{Davydychev:1991va} we write the tensor of rank $2$ as ($[d+]^l=d+2 l, ~d= 4-2\varepsilon$)
\bea
\label{tensor2}
I_{5}^{\mu\, \nu}=
   \sum_{i,j=1}^{4} \, q_i^{\mu}\, q_j^{\nu} \, {\nu}_{ij} \,  \, I_{5,ij}^{[d+]^2} -\frac{1}{2}
   \, g^{\mu \nu}  \, I_{5}^{[d+]} .
\eea
Inserting
\bea
\frac{1}{2} g^{\mu\, \nu}=\sum_{i,j=1}^{5} {\frac{{i\choose j}_5}{\left(  \right)_5}} \, q_i^{\mu}\, q_j^{\nu} 
\label{gmunu}
\eea
and using the recursion relation
\bea
\label{A522}
{\nu}_{ij} I_{5,ij}^{[d+]^2}&=&-\frac{{0\choose j}_5}{\left(  \right)_5} I_{5,i}^{[d+]} +
 \sum_{s=1,s \ne i}^{5} \frac{{s\choose j}_5}{\left(  \right)_5} I_{4,i}^{[d+],s} +
  {\frac{{i\choose j}_5}{\left(  \right)_5}} I_{5}^{[d+]} ,
\eea
we see that the $g^{\mu \nu}$ terms in \eqref{tensor2} cancel with the result
% \bea
% I_{5}^{\mu\, \nu}=\sum_{i,j=1}^{4}  q_i^{\mu}\, q_j^{\nu} \left\{-\frac{{0\choose j}_5}{\left(  \right)_5}I_{5,i}^{[d+]} +
% \sum_{s=1,s \ne i}^{5}  \frac{{s\choose j}_5}{\left(  \right)_5} I_{4,i}^{[d+],s} \right\},
% \eea
% with
% \bea
% I_{5,i}^{[d+]}=-\frac{{0\choose i}_5}{\left(  \right)_5} I_5+\sum_{s=1,s \ne i}^{5}  \frac{{s\choose i}_5}{\left(  \right)_5} I_{4}^{s}, 
% ~~~~~~~~~~~~~~
% I_{4,i}^{[d+],s}=-\frac{{0s\choose is}_5}{{s\choose s}_5} I_4^s+\sum_{t=1}^{5}  \frac{{ts\choose is}_5}{{s\choose s}_5} I_3^{st}.
% \eea
% \eqref{tensor2} results in
\bea
\label{twoteetwo}
I_{5}^{\mu\, \nu}=I_{5}^{\mu} Q_0^{\nu}-\sum_{s=1}^{5} \left\{Q_0^{s,\mu}I_4^s-\sum_{t=1}^{5} Q_t^{s,\mu}I_3^{st}\right\}Q_s^{\nu},
\eea
\bea
I_{5}^{\mu}=E Q_0^{\mu}-\sum_{s=1}^{5} I_4^s Q_s^{\mu},~~~~~~~~~~~~~~
E=\frac{1}{{0\choose 0}_5}\sum_{s=1}^{5} {s\choose 0}_5 I_4^s.
\eea
A more {direct approach} is the use of the $5$-point recursion in terms of $4$-point functions \cite{Diakonidis:2009fx},
\bea
{
I_5^{\mu_1  \dots \mu_{R-1} \mu}  =I_5^{\mu_1  \dots \mu_{R-1}} Q_0^{\mu} -  \sum_{s=1}^{5}
I_4^{\mu_1  \dots \mu_{R-1},s } Q_s^{\mu}}.
\label{tensor5general}
\eea
{This formula in general takes into account some cancellations of the $g_{\mu \nu}$}. Only {$g_{\mu \nu}$-contributions
from $4$-points have still to be dealt with - and they are simpler to handle}. This we demonstrate for 
the tensor of degree $3$. As a special case of \eqref{tensor5general} we have
\bea\label{3.138}
I_5^{\mu \nu \lambda}=I_5^{\mu\nu} \cdot Q_0^{\lambda} -\sum_{s=1}^{5} I_4^{\mu\nu, s} \cdot Q_s^{\lambda} .
\eea
The corresponding $4$-point function reads ($q_5=0$):
\bea
\label{TwoTy}
I_4^{\mu \nu ,s}&=&\sum_{i,j=1}^{4} q_i^{\mu} q_j^{\nu} {\nu}_{ij} I_{4,ij}^{[d+]^2,s}-\frac{1}{2} g^{\mu \nu}I_{4}^{[d+],s}, \nn
\\
\label{TwoTyTwo}
{\nu}_{ij} I_{4,ij}^{[d+]^2,s}&=&-\frac{ {0s\choose js}_5}{{s\choose s}_5} I_{4,i}^{[d+],s}+
\frac{ {is\choose js}_5}{{s\choose s}_5}I_{4}^{[d+],s}+
\sum_{t=1}^{5}
\frac{ {ts\choose js}_5}{{s\choose s}_5} I_{3,i}^{[d+],st} ,
\eea
and with
\bea
\frac{ {is\choose js}_5}{{s\choose s}_5}&=&\frac{ {i\choose j}_5}{{\left( \right)}_5}
-\frac{ {s\choose i}_5 {s\choose j}_5 }{{\left( \right)}_5 {s\choose s}_5} 
\label{trick}
\eea
we again observe the possibility to cancel $g^{\mu \nu}$ with the result
\bea
I_4^{\mu \nu,s}&=&Q_0^{s,\mu} Q_0^{s,\nu} I_4^s-\frac{{\left( \right)}_5}{{s\choose s}_5}Q_s^{\mu} Q_s^{\nu} I_4^{[d+],s}
-\sum_{t=1}^5 \left\{Q_t^{s,\mu} Q_0^{s,\nu} I_3^{st}+Q_t^{s,\nu}  I_3^{\mu , st} \right\}, \nn \\
I_3^{\mu ,st}&=& - \sum_{i=1}^4 q_i^{\mu} I_{3,i}^{[d+],st} = 
Q_0^{st,\mu} I_3^{st}-\sum_{u=1}^5 Q_u^{st,\mu} I_2^{stu}.
\label{I4munu}
\eea

% A remark concerning the {$4$-point function}, only {3} of the 4 vectors $q_i, i=1, \dots 4$ \red{independent}:
% 
% \bea
% g_{\mu \nu} \rightarrow
% G^{\mu \nu}=g^{\mu \nu}-2 \sum_{i,j=1}^4 q_i^{\mu} q_j^{\nu}
% \frac{{i\choose j}}{{\left( \right)_5}}
% =\frac{8 v^{\mu} v^{\nu}}{{\left( \right)_5}} ,
% \label{Gg}
% \eea
% with
% \bea\label{eq-wa-6}
% v^{\mu}={\eps}^{\mu \lambda \rho \sigma} (q_1-q_4)_{\lambda}(q_2-q_4)_{\rho}(q_3-q_4)_{\sigma}.
% \eea
% Assume $q_4=0$ $\rightarrow$ ($q_i \cdot v$)~=~0 , i.e. $q_{i \mu} G^{\mu \nu}~=~0$ : \red{effectively vanishing !}

%======================================================================
\section{\label{5to4}Contracting the tensor integrals}
%======================================================================
Scalar expressions, contracting with chords, are
\bea
\label{defini1}
q_{i_1\mu_1}\cdots q_{i_R\mu_R} ~~I_{5}^{\mu_1\cdots\mu_R}& =&
~\int\frac{ d^d k~}{i {\pi}^{d/2}}~\frac{\prod_{r=1}^{R} (q_{i_r} \cdot k)}{\prod_{j=1}^{5}c_j},  \\
\label{defini2}
g_{\mu_1,\mu_2}
q_{i_1\mu_3}\cdots q_{i_R\mu_R} ~~I_{5}^{\mu_1\cdots\mu_R} &=&
~\int\frac{k^2 d^d k~}{i {\pi}^{d/2}}~\frac{\prod_{r=3}^{R} (q_{i_r} \cdot k)}{\prod_{j=1}^{5}c_j}, 
\eea
etc.
Eqns. \eqref{defini1} and \eqref{defini2} define the contraction of all tensor indices
with chords and the direct contraction of two tensor indices, respectively.
 These are obtained
in realistic matrix element calculations
 by constructing projection operators or
by  constructing scalar differential cross sections 
(Born $\times$ 1-loop) before loop integration.
 As a result of these contractions
the $1 / \left( \right)_5$  cancels already.

To begin with, we have a look at %the scalar obtained from the tensor of degree $2$
\bea
q_{a \mu} q_{b \nu} I_5^{\mu \nu} =(q_a \cdot I_5) (q_b \cdot Q_0) 
 -\sum_{s=1}^{5} \left\{ (q_a \cdot Q_0^s) I_4^s
-\sum_{t=1}^{5} (q_a \cdot Q_t^s) I_3^{st} \right\} (q_b \cdot Q_s).
\eea
 For $q_n=0,~~ a=1, \dots , n-1,~~s=1, \dots n$
\bea
\label{Zcalar1}
(q_a \cdot Q_0) &=&\sum_{j=1}^{n-1} (q_a  \cdot q_j) \frac{{0\choose j}_n}
{{\left(\right)}_n}=-\frac{1}{2}\left(
 Y_{an}-Y_{nn} \right), \\
(q_a \cdot Q_s) &=&\sum_{j=1}^{n-1} (q_a  \cdot q_j) \frac{{s\choose j}_n}
{{\left( \right)}_n}=~~~\frac{1}{2}
\left({\delta}_{as}-{\delta}_{ns}\right) ,
\label{Zcalar2}
\eea
and 
\bea
(q_a \cdot I_5)= E (q_a \cdot Q_0)-\sum_{s=1}^5 I_4^s (q_a \cdot Q_s).
\eea
Further
\bea
\label{contrs}
(q_a \cdot Q_0^s) = \frac{1}{{s\choose s}_5} \Sigma_a^{2,s}, ~~~ 
(q_a \cdot Q_t^s) = \frac{1}{{s\choose s}_5} \Sigma_a^{1,st},
\eea
where the sums $\Sigma_a^{2,s}$ and $\Sigma_a^{1,st}$ are given in \cite{Fleischer:2011nt}.
Both are linear combinations of ${s\choose s}_5$ and Kronecker-$\delta$'s, $\left({\delta}_{as}-{\delta}_{5s}\right) $.
Indeed the fact that there is no inverse $\left( \right)_5$ anymore is due to relations \eqref{Zcalar1} and
\eqref{Zcalar2}. The second scalar which can be 
constructed from the tensor of degree $2$ is $g_{\mu \nu} I_5^{\mu \nu}$. Due to
\eqref{twoteetwo} we need to evaluate the following scalar products:
 \bea
(Q_0 \cdot Q_0)&=&\frac{1}{2}\left[\frac{{0\choose 0}_5}{\left(  \right)_5}+Y_{55} \right], \nn \\
(Q_0 \cdot Q_s)&=&\frac{1}{2}\left[\frac{{s\choose 0}_5}{\left(  \right)_5}-{\delta}_{s5} \right], \nn \\
(Q_0^s \cdot Q_s)&=&-\frac{1}{2}{\delta}_{s5}, \nn \\
(Q_t^s \cdot Q_s)&=&~~~0.
\label{scalpr}
\eea
In this case the terms with ${1} / {\left(  \right)_5}$ cancel and, not surprisingly, the result finally is
\bea
\label{Nosurprise}
g_{\mu \nu} I_5^{\mu \nu}=\frac{Y_{55}}{2} E + I_4^5.
\eea
To calculate $g_{\mu \nu} I_5^{\mu \nu \lambda}$ we need $g_{\mu \nu} I_4^{\mu \nu, s }$ and thus
 further scalar products, see \eqref{I4munu}:
\bea
(Q_0^s \cdot Q_0^s)&=&\frac{1}{2 {s\choose s}_5}\left[{0s\choose 0s}_5+2 {s\choose 0}_5 {\delta}_{s5}\right]+
                       \frac{1}{2} Y_{55} , \nn \\               
(Q_s \cdot Q_s)&=&\frac{1}{2}\frac{{s\choose s}_5}{\left(  \right)_5}, \nn \\
(Q_t^s \cdot Q_0^s)&=&\frac{1}{2 {s\choose s}_5}\left[{ts\choose 0s}_5-{s\choose s}_5 {\delta}_{t5}+{s\choose t}_5 {\delta}_{s5}\right],
\nn \\
(Q_t^s \cdot Q_0^{st})&=&\frac{1}{2 {s\choose s}_5}\left[~~~~~~~~~~~-{s\choose s}_5 {\delta}_{t5}+{s\choose t}_5 {\delta}_{s5}\right],
\nn \\
(Q_t^s \cdot Q_u^{st})&=&0, 
\eea
which yields
\bea
\label{nottriv}
g_{\mu \nu} I_4^{\mu \nu, s }=
\frac{ Y_{55}}{2} I_4^s + I_3^{s5} + \frac{{\delta}_{s5}}{{s\choose s}_5} \left[{s\choose 0}_5 I_4^s - \sum_{t=1}^5 {s\choose t}_5 I_3^{st}\right] 
\eea
and finally
\bea
\label{twocon}
g_{\mu \nu} I_5^{\mu \nu \lambda}=-\frac{Y_{55}}{2} \sum_{s=1}^5 I_4^s~ Q_s^{0,\lambda} + I_4^5 Q_0^{5,\lambda}-\sum_{t=1}^4 I_3^{5t} Q_t^{5,\lambda}.
\eea
It is remarkable that \eqref{nottriv} is trivial again for $s \ne 5$. For $s=5$, however, the standard cancelation
of propagators does not work and for this case \eqref{nottriv} is indeed a useful result.
For further contraction of \eqref{twocon} with a vector $q_{\lambda}$ again \eqref{contrs} can be applied.
%======================================================================
\section{\label{Avoi}Avoiding inverse $4$-point Gram determinants}
%======================================================================
While in the above approach of taking scalar products of the tensors with chords the inverse
$\left( \right)_5$ Gram determinant cancels already, there still remains the inverse ${s\choose s}_5$
sub-Gram determinant of the $4$-point functions. Therefore we have to choose a different approach for 
the case if the latter becomes small. This approach consists in avoiding the inverse $\left( \right)_5$
already in the $5$-point tensors from the very beginning and keeping only $4$-point integrals in higher 
dimensions (i.e. integrals with only powers 1 of the scalar propagators), which
for small ${s\choose s}_5$ should be evaluated in a different manner than by standard recursion,
see \cite{Fleischer:2010sq}. If one does not want to reintroduce the inverse $\left( \right)_5$
to see the cancellation of the $g^{\mu \nu}$, one can explicitely see its cancelation also after
taking contractions.
For the tensor of degree $2$ we refer to \cite{Fleischer:2011nt} for the contraction with two chords.
For the self-contraction of the tensor indices no simpler result than \eqref{Nosurprise} can be achieved 
anyway. For the tensor of degree $3$ we present new results for the contraction with three chords
and a self-contraction.

\vspace{0.5cm}

The tensor can be written as follows (see \cite{Fleischer:2010sq} (4.35)-(4.37)):
\begin{eqnarray}
I_{5}^{\mu\, \nu\, \lambda}&&= \sum_{i,j,k=1}^{5} \, q_i^{\mu}\, q_j^{\nu} \, q_k^{\lambda}
E_{ijk}+\sum_{k=1}^5 g^{[\mu \nu} q_k^{\lambda]} E_{00k},
\label{Exyz0}
\end{eqnarray}
with
\bea
\label{Exyz1}
E_{00k} 
&=& \sum_{s=1}^5 \frac{1}{{0\choose 0}_5}  \left[\frac{1}{2} {0s\choose 0k}_5 I_4^{[d+],s}- \frac{d-1}{3} {s\choose k}_5 I_4^{[d+]^2,s} \right]  ,
\\
E_{ijk} 
&=&-  \sum_{s=1}^5\frac{1}{{0\choose 0}_5}  \left\{ \left[{0j\choose sk}_5 I_{4,i}^{[d+]^2,s}+
(i \leftrightarrow j)\right]+{0s\choose 0k}_5 {\nu}_{ij} I_{4,ij}^{[d+]^2,s} \right\}.
\label{Exyz2}
\eea

\vspace{0.5cm}

Contraction of the tensor with three chords yields:
\bea
\label{Proj3}
&&q_{a \mu} q_{b \nu} q_{c \lambda}I_{5}^{\mu\, \nu\, \lambda}= 
\sum_{i,j,k=1}^4 
(q_a \cdot q_i) (q_b \cdot q_j) (q_c \cdot q_k)E_{ijk}~~~~~~~~~~~~~~~~~~~~~~~~~~~~~ \nn \\ 
&&~~~~+~ \sum_{k=1}^4 \left[(q_a \cdot q_b) (q_c \cdot q_k) +(q_a \cdot q_c) (q_b \cdot q_k)+(q_b \cdot q_c) (q_a \cdot q_k)\right] E_{00k} . 
\eea

\vspace{0.5cm}

Introducing ${\Yps}_a =Y_{a5}-Y_{55}$, ${\Dps}_a^s ={\delta}_{as}-{\delta}_{5s}$ and the following kinematical
objects,
\bea
\label{P1I4d}
P_{I4}&=&\frac{1}{8}\frac{1}{{0s\choose 0s}_5} \left\{ {s\choose s}_5 \left[ {s\choose 0}_5 {\Yps}_a {\Yps}_b {\Yps}_c+
{0\choose 0}_5 \left( {\Yps}_a {\Yps}_b {\Dps}_c^s +  {\Yps}_a {\Yps}_c {\Dps}_b^s +  {\Yps}_b {\Yps}_c {\Dps}_a^s \right) \right]
\right. \nn \\ 
&& \left. +{0\choose 0}_5 \left[ {s\choose 0}_5 \left( {\Yps}_a {\Dps}_b^s {\Dps}_c^s + {\Yps}_b {\Dps}_a^s {\Dps}_c^s +
{\Yps}_c {\Dps}_a^s {\Dps}_b^s \right) + {0\choose 0}_5 {\Dps}_a^s {\Dps}_b^s {\Dps}_c^s \right] 
\right\}, \\
\label{P3I4d}
P_{Z4}&=&P_{I4}-\frac{1}{12} {\left( \right)}_5 \left\{{\Yps}_a {\Yps}_b {\Dps}_c^s +{\Yps}_a {\Yps}_c {\Dps}_b^s +
{\Yps}_b {\Yps}_c {\Dps}_a^s \right\},
\\
% ~~~{\rm and} \\
P_{I3}&=&\frac{1}{24}\frac{{\left( \right)}_5 }{{0s\choose 0s}_5}\left\{
\left[ {\Yps}_a {\Dps}_b^s+{\Yps}_b {\Dps}_a^s \right] \left[{0s\choose 0s}_5 {\Dps}_c^t-{0s\choose 0t}_5 {\Dps}_c^s \right]+
(a \leftrightarrow c) + (b \leftrightarrow c) 
\right\} , 
\label{C3remain}
\nn \\
\eea

\vspace{0.5cm}

we can write

%%%%%%%%%%%%%%%%%%%%%%%%%%%%%%%%%%%%%%%%%%%%%%%%%%
\newpage

\bea
q_{a \mu} q_{b \nu} q_{c \lambda} I_{5}^{\mu\, \nu\, \lambda}
=
&& 
\frac{d-2}{8} \frac{{\left(\right)}_5}{{0s\choose 0s}_5}
({\delta}_{ab}{\delta}_{ac}{\delta}_{as}-{\delta}_{5s}) (d-1)I_{4}^{[d+]^2,s} -
\frac{1}{{0\choose 0}_5} \left\{ P_{I4}~ I_4^{[d+],s} 
\nn 
\right. 
\\
&&
\left. -P_{Z4} ~Z_4^{[d+],s}+P_{I3}~ I_3^{[d+],st} +\frac{1}{3}\left[ {\Sigma}_{c}^{1,s} R_{ab} + {\Sigma}_{b}^{1,s} R_{ac} +{\Sigma}_{a}^{1,s} R_{cb} \right] \right\} 
\nn 
\\
&&
+ F_{abc}^s, 
\eea
symmetric in the indices $a,b,c$ and summation over $s,t$ assumed. Here, the
\bea
R_{ab} =&&
-~\frac{1}{{0s\choose 0s}_5}
%\sum_{t=1}^{5}
\left\{\frac{1}{{0s\choose 0s}_5}{\Sigma}^{2,s}_{b}{\Sigma}^{2,st}_{a}(d-2)I_3^{[d+],st} \right. \nn \\
&&+~ \left.
\frac{1}{{0st\choose 0st}_5}{\Sigma}^{2,st}_{b} 
\left[{\Sigma}^{3,st}_{a}(d-2)I_3^{[d+],st}-\sum_{u=1}^{5}{\Sigma}^{2,stu}_{a}I_2^{stu} \right] \right\}
\eea
contains only $3$-point functions and no inverse ${s\choose s}_5$.
Further,
\bea
F_{abc}^s=-\frac{1}{24} \frac{{\left(\right)}_5}{{0s\choose 0s}_5}
\frac{{s\choose 0}_5 }{{0\choose 0}_5 } \left[{\Yps}_c {\Dps}_a^s {\Dps}_b^s+{\Yps}_b {\Dps}_a^s {\Dps}_c^s+{\Yps}_a {\Dps}_b^s {\Dps}_c^s   \right]
\label{term2}
\eea
is a rational term obtained from an $\eps$-expansion. The fact that no scalar products from \eqref{Proj3} remain
demonstrates that the $g^{\mu \nu}$ term has canceled.

\vspace{0.5cm}

For the selfcontracted tensor we obtain
\bea
\label{resultkkv}
&&
q_{a \lambda}I_{5,\mu}^{~~~~\mu\, \lambda}
= 
-\frac{1}{{0\choose 0}_5}\sum_{s=1}^5\left\{{\Sigma}_a^{1,s} \left[1+\frac{1}{2} \frac{1}{{0s\choose 0s}_5}
\left({s\choose s}_5 Y_{55}+ 2 {s\choose 0}_5 {\delta}_{s5} \right)\right] \right. 
\nn 
\\
&& 
\left. ~~~~~~~~~~~~~~~~~~~~~~~~~~~~~~~~~~+ \left( \right)_5
\left(Y_{a5}-Y_{55}\right) {\delta}_{s5}  \right\} I_4^{[d+],s} \nn \\
&&-\frac{1}{{0\choose 0}_5}\sum_{s,t=1}^5 {\Sigma}_a^{1,s} \frac{{ts\choose 0s}_5}{{0s\choose 0s}_5}
\frac{Y_{55}}{{0st\choose 0st}_5} \left[\frac{d-2}{2}{st\choose st}_5 I_3^{[d+],st}+\frac{1}{2}
\sum_{u=1}^5 {st0\choose stu}_5 I_2^{stu} \right]\nn \\
&&+\frac{{\delta}_{s5}}{{0\choose 0}_5}\sum_{t=1}^5 {\Sigma}_a^{1,s}\frac{1}{{0s\choose 0s}_5
{0st\choose 0st}_5}\left\{\left[{0s\choose 0t}_5{st\choose st}_5~~-{ts\choose 0s}_5{st\choose 0t}_5
~~\right]\frac{d-2}{2}I_3^{[d+],st} \nn \right. 
\\ &&
\left.~~~~~~~~~~~~~~~~~~~~~~~~~~~~~~~~~~+
\sum_{u=1}^5\left[{0s\choose 0t}_5{st0\choose stu}_5-{ts\choose 0s}_5{st0\choose 0tu}_5\right] \frac{1}{2}I_2^{stu} \right\}\nn \\
&&-\frac{{\delta}_{t5}}{{0\choose 0}_5}\sum_{s=1}^5 {\Sigma}_a^{1,s}\frac{1}{{0s\choose 0s}_5
{0st\choose 0st}_5}\left\{\left[{0s\choose 0s}_5{st\choose st}_5~~+{ts\choose 0s}_5{ts\choose 0s}_5
~~\right]\frac{d-2}{2}I_3^{[d+],st} \nn \right. 
\\ &&
\left.~~~~~~~~~~~~~~~~~~~~~~~~~~~~~~~~~~+
\sum_{u=1}^5\left[{0s\choose 0s}_5{st0\choose stu}_5+{ts\choose 0s}_5{ts0\choose 0su}_5\right] \frac{1}{2}I_2^{stu} \right\}\nn \\
&&+\frac{1}{{0\choose 0}_5}\sum_{s,t=1}^5 {\Sigma}_a^{1,s}\frac{1}{{0s\choose 0s}_5} \left\{
\left[{s\choose t}_5{\delta}_{s5}-{s\choose s}_5{\delta}_{t5}\right]\frac{d-2}{2}I_3^{[d+],st} -
{ts\choose 0s}_5\frac{1}{2}I_2^{st5} \right\}\nn \\
&&+\frac{1}{{0\choose 0}_5}\sum_{s,t=1}^5
\left[{t\choose 0}_5 \left({\delta}_{as}-{\delta}_{5s} \right)-
{s\choose 0}_5 \left({\delta}_{at}-{\delta}_{5t} \right) \right] I_3^{[d+],st}.
\eea
The first two lines of \eqref{resultkkv} contain complete double sums while the remaining terms
contribute only for specific values of $s,t,u$.

% %=================================================

\section*{Acknowledgements}
J.F. thanks DESY for kind hospitality.
Work is supported in part by Sonderforschungsbereich/Trans\-re\-gio SFB/TRR 9 of DFG
``Com\-pu\-ter\-ge\-st\"utz\-te Theoretische Teil\-chen\-phy\-sik"
and European Initial Training Network LHCPHENOnet PITN-GA-2010-264564.
%%%%%%%%%%%%%%%%%%%%%%%%%%%%%%%%%%%%%%%%%%%%%%%%%%%%%%%%%%%%%%%%%%%%%%%%%%

\end{document}